\documentclass[10pt, final, twocolumn, twoside, romanappendices]{IEEEtran}

\usepackage{showtags} 
\usepackage{graphicx,courier,amssymb,amsmath,times}
\usepackage{epsfig}
\usepackage{bm,graphicx,psfrag}
\usepackage{amsfonts}
\usepackage[latin1]{inputenc}
\usepackage{booktabs}
\usepackage{algorithmic}
\usepackage{caption}
\usepackage{cite}
\usepackage{hyperref}
\usepackage[usenames]{color}
\usepackage[nolist]{acronym}
\usepackage{currfile} %\currfilename is the current file name
\usepackage{File/slashbox}
\usepackage[dvipsnames]{xcolor}
\usepackage{tikz}
\usepackage{pgfplots}
\usepackage{File/Symbols_QECC}
\usepackage[ruled,vlined,noresetcount]{algorithm2e}
\usepackage{physics}
\usepackage{mathtools}

\usepackage{amsthm}

\input{File/winsnotation.sty}

\algsetup{linenosize=\scriptsize}

\newtheorem{theorem}{Theorem} %[section]

\usepackage[yyyymmdd,hhmmss]{datetime} % gives time of the compiling
\newdateformat{monthyeardate}{\monthname[\THEMONTH] \THEDAY, \THEYEAR} % specifies the format of the date
\newcommand{\Syndrome}[0]{\V{s}}

\pgfplotsset{compat=1.16}
\begin{document}

%\title{Design of Short Codes for Asymmetric Quantum Channels}
\title{Short Codes for Quantum Channels with One Prevalent Pauli Error Type}
\author{Marco~Chiani,~\IEEEmembership{Fellow,~IEEE}, and Lorenzo Valentini 
\\
\thanks{Manuscript received February 11, 2020; revised June 1, 2020;  accepted July 20, 2020. Date of publication MONTH DD, 2020; date of current version MONTH DD, 2020. This work was supported in part by the Italian Ministry for Education, University and Research under the program Dipartimenti di Eccellenza (2018-2022). This paper was presented in part at the Quantum Computing Workshop, 2020 International Conference on Computational Science (ICCS).}
\thanks{The authors are with the Department of Electrical, Electronic, and Information Engineering ``Guglielmo Marconi'' and CNIT, University of Bologna, 40136 Bologna, Italy (e-mail: \texttt{\{marco.chiani, lorenzo.valentini13\}@unibo.it}). }
\thanks{Digital Object Identifier 10.1109/JSAIT.2020.XXXXXXX}
}
%

%\markboth{Vol.~X, No.~Y, Month~2020}{Asymmetric Short Codes}
\markboth{IEEE JOURNAL ON SELECTED AREAS IN INFORMATION THEORY, VOL. X, NO. X, AUGUST 2020}{Chiani, Valentini: Short Codes for Quantum Channels with One Prevalent Pauli Error Type}

\maketitle

\begin{acronym}
% usage: \ac{SW}, \acp{SW} for plurals \acf{SW} Use the full name of the acronym.
%\acs{SW}Use the acronym, even before the first corresponding \ac command
%\acl{acronym}Expand the acronym without using the acronym itself.
\small
\acro{CWER}{Codeword Error Rate}
\acro{CWEP}{Codeword Error Probability}
\acro{AcR}{autocorrelation receiver}
\acro{ACF}{autocorrelation function}
\acro{ADC}{analog-to-digital converter}
\acro{AWGN}{additive white Gaussian noise}
\acro{BCH}{Bose Chaudhuri Hocquenghem}
\acro{BEP}{bit error probability}
\acro{BFC}{block fading channel}
\acro{BPAM}{binary pulse amplitude modulation}
\acro{BPPM}{binary pulse position modulation}
\acro{BPSK}{binary phase shift keying}
\acro{BPZF}{bandpass zonal filter}
\acro{CD}{cooperative diversity}
\acro{CDF}{cumulative distribution function}
\acro{CCDF}{complementary cumulative distribution function}
\acro{CDMA}{code division multiple access}
\acro{c.d.f.}{cumulative distribution function}
\acro{c.c.d.f.}{complementary cumulative distribution function}
\acro{ch.f.}{characteristic function}
\acro{CIR}{channel impulse response}
\acro{CR}{cognitive radio}
\acro{CSI}{channel state information}
\acro{DAA}{detect and avoid}
\acro{DAB}{digital audio broadcasting}
\acro{DS}{direct sequence}
\acro{DS-SS}{direct-sequence spread-spectrum}
\acro{DTR}{differential transmitted-reference}
\acro{DVB-T}{digital video broadcasting\,--\,terrestrial}
\acro{DVB-H}{digital video broadcasting\,--\,handheld}
\acro{ECC}{European Community Commission}
\acro{ELP}{equivalent low-pass}
\acro{FCC}{Federal Communications Commission}
\acro{FEC}{forward error correction}
\acro{FFT}{fast Fourier transform}
\acro{FH}{frequency-hopping}
\acro{FH-SS}{frequency-hopping spread-spectrum}
\acro{GA}{Gaussian approximation}
\acro{GPS}{Global Positioning System}
\acro{HAP}{high altitude platform}
\acro{i.i.d.}{independent, identically distributed}
\acro{IFFT}{inverse fast Fourier transform}
\acro{IR}{impulse radio}
\acro{ISI}{intersymbol interference}
\acro{LEO}{low earth orbit}
\acro{LOS}{line-of-sight}
\acro{BSC}{binary symmetric channel}
\acro{MB}{multiband}
\acro{MC}{multicarrier}
\acro{MF}{matched filter}
\acro{m.g.f.}{moment generating function}
\acro{MI}{mutual information}
\acro{MIMO}{multiple-input multiple-output}
\acro{MISO}{multiple-input single-output}
\acro{MRC}{maximal ratio combiner}
\acro{MMSE}{minimum mean-square error}
\acro{M-QAM}[$M$-QAM]{$M$-ary quadrature amplitude modulation}
\acro{M-PSK}[${M}$-PSK]{$M$-ary phase shift keying}
\acro{MUI}{multi-user interference}
\acro{NB}{narrowband}
\acro{NBI}{narrowband interference}
\acro{NLOS}{non-line-of-sight}
\acro{NTIA}{National Telecommunications and Information Administration}
\acro{LDPC}{low density parity check}
\acro{OC}{optimum combining}
\acro{OFDM}{orthogonal frequency-division multiplexing}
\acro{p.d.f.}{probability distribution function}
\acro{PAM}{pulse amplitude modulation}
\acro{PAR}{peak-to-average ratio}
\acro{PDP}{power dispersion profile}
\acro{p.m.f.}{probability mass function}
\acro{PN}{pseudo-noise}
\acro{PPM}{pulse position modulation}
\acro{PRake}{Partial Rake}
\acro{PSD}{power spectral density}
\acro{PSK}{phase shift keying}
\acro{QAM}{quadrature amplitude modulation}
\acro{QPSK}{quadrature phase shift keying}
\acro{8-PSK}[$8$-PSK]{$8$-phase shift keying}
\acro{r.v.}{random variable}
\acro{R.V.}{random vector}
\acro{SEP}{symbol error probability}
\acro{SIMO}{single-input multiple-output}
\acro{SIR}{signal-to-interference ratio}
\acro{SISO}{single-input single-output}
\acro{SINR}{signal-to-interference plus noise ratio}
\acro{SNR}{signal-to-noise ratio}
\acro{SS}{spread spectrum}
\acro{TH}{time-hopping}
\acro{ToA}{time-of-arrival}
\acro{TR}{transmitted-reference}
\acro{UAV}{unmanned aerial vehicle}
\acro{UWB}{ultrawide band}
\acro{UWBcap}[UWB]{Ultrawide band}
\acro{WLAN}{wireless local area network}
\acro{WMAN}{wireless metropolitan area network}
\acro{WPAN}{wireless personal area network}
\acro{WSN}{wireless sensor network}
\acro{WSS}{wide-sense stationary}

\acro{SW}{sync word}
\acro{FS}{frame synchronization}
\acro{FSsmall}[FS]{frame synchronization}
\acro{BSC}{binary symmetric channels}
\acro{LRT}{likelihood ratio test}
\acro{GLRT}{generalized likelihood ratio test}
\acro{LLRT}{log-likelihood ratio test}
\acro{LLR}{log-likelihood ratio}
\acro{$P_{EM}$}{probability of emulation, or false alarm}
\acro{$P_{MD}$}{probability of missed detection}
\acro{ROC}{receiver operating characteristic}
\acro{AUB}{asymptotic union bound}
\acro{RDL}{"random data limit"}
\acro{PSEP}{pairwise synchronization error probability}

\acro{SCM}{sample covariance matrix}

\acro{QECC}{quantum error-correcting code}
\acro{CECC}{classical error-correcting code}
\acro{DMC}{discrete memoryless channel}
\acro{PSC}{piggyback syndrome channel}
\acro{QN}{quantum network}
\acro{IP}{Internet Protocol}
\acro{CSS}{Calderbank Shor Steane}
\acro{QHB}{quantum Hamming bound}
\acro{GQHB}{generalized quantum Hamming bound}

\end{acronym}

\begin{abstract}  
One of the main problems in quantum information systems is the presence of errors due to noise, and for this reason \acp{QECC} play a key role. While most of the known codes are designed for correcting generic errors, i.e., errors represented by arbitrary combinations of Pauli $\PauliX, \PauliY$ and $\PauliZ$ operators, in this paper we investigate the design of stabilizer \ac{QECC} able to correct a given number $\eg$ of generic Pauli errors, plus $\eZ$ Pauli errors of a specified type, e.g., $\PauliZ$ errors. These codes can be of interest when the quantum channel is asymmetric in that some types of error occur more frequently than others. We first derive a generalized quantum Hamming bound for such codes, then propose a design methodology based on syndrome assignments.   
For example, we found a $[[9,1]]$ quantum error-correcting code able to correct up to one generic qubit error plus one $\PauliZ$ error in arbitrary positions. This, according to the generalized quantum Hamming bound, is the shortest code with the specified error correction capability. Finally, we evaluate analytically the performance of the new codes over asymmetric channels.  
\end{abstract}


\section{Introduction}
\label{sec:intro}

The possibility to exploit the unique features of quantum mechanics is paving the way to new approaches for acquiring, processing and transmitting information  \cite{Zol:05,kim:08,Weh:18,NAP:19,OSW18}. 
In this regard, one of the main problems is the noise caused by unwanted interaction of the quantum information with the environment. 
Error correction techniques are therefore essential for quantum computation, quantum memories and quantum communication systems \cite{Kni:1997,Ter:15,Mun:12,Mur:16}.  
Compared to the classical case, quantum error correction is made more difficult by the laws of quantum mechanics which imply that qubits cannot be copied or measured without perturbing state superposition \cite{Woo:1982}. 
Moreover, there is continuum of errors that could occur on a qubit.  
However, it has been shown that in order to correct an arbitrary qubit error it is sufficient to consider error correction on the discrete set of Pauli operators, i.e., the bit flip $\PauliX$, phase flip $\PauliZ$, and combined bit-phase flip $\PauliY$ \cite{BenDivSmoWoo:96,Kni:1997,NieChu:2010, Got:2009}. 
Hence, we can consider in general a channel introducing qubit errors $\PauliX$, $\PauliY$, and $\PauliZ$ with probabilities $\px$, $\py$, and $\pz$, respectively, and leaving the qubit intact with probability $1-\rho$, where $\rho = \px + \py + \pz$. 
A special case of this model is the so-called depolarizing channel for which $\px = \py = \pz = \rho/3$. Quantum error-correcting codes for this channel are thus naturally designed to protect against equiprobable Pauli errors 
\cite{Sho:1995, Ste:1996, Laf:1996}.

However, not all channels exhibit this symmetric behaviour of Pauli errors  as, in some situations, some types of error are more likely than others \cite{Iof:2007}. In fact, depending on the technology adopted for the   system implementation, the different types of Pauli error can have quite different probabilities of occurrence, leading to asymmetric quantum channels \cite{Sar:2009,Gyo:2018,Lay:2020}. 

Owing to this considerations, it can be useful to investigate the design of quantum codes with error correction capabilities tailored to specific channel models. For example, codes for the amplitude damping channel have been proposed in \cite{Fle:08,Fle:08b,Lan:07,Leu:97,Sho:11,Jac:16}, 
while quantum error-correcting codes for more general asymmetric channels are investigated in \cite{Eva:2007,Iof:2007,Sar:2009,Gyo:2018}. 
In particular, asymmetric \ac{CSS} codes, where the two constituent classical parity check matrices are chosen with different error correction capability (e.g., \ac{BCH} codes for $\PauliX$ errors and \ac{LDPC} codes for $\PauliZ$ errors),  are investigated in \cite{Iof:2007,Sar:2009}. 
Inherent to the \ac{CSS} construction there are two distinct error correction capabilities for the $\PauliX$ and the $\PauliZ$ errors; the resulting  asymmetric codes, denoted as $[[n,k,d_{\mathrm{X}}/d_{\mathrm{Z}}]]$, can correct up to $t_{\mathrm{X}} = \lfloor(d_{\mathrm{X}}-1)/2\rfloor$ Pauli $\PauliX$ errors and $t_{\mathrm{Z}} = \lfloor(d_{\mathrm{Z}}-1)/2\rfloor$ Pauli $\PauliZ$ errors per codeword. 
Due to the possibility of employing tools from classical error correction, many works have been focused on asymmetric codes based on the \ac{CSS} construction, which, however, may not lead to the shortest codes (e.g., for the symmetric channel compare the $[[7,1]]$ \ac{CSS} code with the shortest $[[5,1]]$ code \cite{Ste:1996,Laf:1996}). 

In this paper we consider the Pauli-twirled asymmetric channel associated to the combination of  amplitude damping and dephasing channels \cite{Sar:2009}. This model has $\px = \py$ and $\pz = A\rho/(A+2)$, where  $\rho$ is the error probability, and the asymmetry is accounted for by the parameter $A = \pz/\px$. This parameter is a function of the relaxation time, $T_{\mathrm 1}$, and the dephasing time, $T_{\mathrm 2}$,  which are in general different, leading to $A > 1$ \cite{Eva:2007,Gyo:2018}. 
For this channel we study stabilizer codes able to correct a given number $\eg$ of generic Pauli  errors, plus a number $\eZ$ of Pauli errors of a specified type (e.g., $\PauliZ$ errors). We denote these as asymmetric $[[n,k]]$ codes with correction capability $(\eg, \eZ)$. Since we are targeting the shortest codes we do not constrain the design to \ac{CSS} codes.

Specifically, we first derive a generalized version of the quantum Hamming bound, which was originally developed for codes able to correct up to a number $t$ of generic errors. The new  generalized bound is valid also for codes with asymmetric error correction capability $(\eg, \eZ)$. 
Then, we construct, by a procedure based on syndrome assignment, a $[[9,1]]$ code with ($\eg=1$, $\eZ=1$) which, according to the new quantum Hamming bound, is the shortest possible code. 
We extend the construction method to the class of $[[n,1]]$ codes with $\eg = 1$ and arbitrary $\eZ$, and provide as an example a $[[13,1]]$ code with ($\eg=1$, $\eZ=2$). Finally, we analytically compare the error correction capability of the new and of previously known codes, over asymmetric channels. 

\subsection{Notation}
Throughout the paper we will use the following notation. A qubit is an element of the two-dimensional Hilbert space $\mathcal{H}^{2}$, 
with basis $\ket{0}$ and $\ket{1}$ \cite{NieChu:2010}. 
An $n$-tuple of qubits ($n$ qubits) is an element of the $2^n$-dimensional Hilbert space, $\mathcal{H}^{2^n}$\!, with basis composed by all possible tensor products $\ket{i_1} \ket{i_2} \cdots \ket{i_n}$, with $i_j \in \{0,1\}, 1\le j\le n$.  
The Pauli operators, denoted as $\M{I}, \M{X}, \M{Z}$, and $\M{Y}$, are defined by  $\M{I}\ket{a}=\ket{a}$, $\M{X}\ket{a}=\ket{a\oplus 1}$, $\M{Z}\ket{a}=(-1)^a\ket{a}$, and $\M{Y}\ket{a}=i(-1)^a\ket{a\oplus 1}$ for $a \in \lbrace0,1\rbrace$. These operators either commute or anticommute. 
With $[[n,k]]$ we indicate a \ac{QECC} that encodes $k$ data qubits $\ket{\varphi}$ into a codeword of $n$ qubits $\ket{\psi}$.  
We use the stabilizer formalism, where a stabilizer code $\mathcal{C}$ is generated by $n-k$ independent and commuting operators $\M{G}_i \in \mathcal{G}_n$, called generators \cite{Got:96,gottesman2009introduction,NieChu:2010}.  
The  code $\mathcal{C}$ is the set of quantum states $\ket{\psi}$ satisfying  
$\M{G}_i \ket{\psi}=\ket{\psi} \,,  i=1, 2, \ldots, n-k\,.
$ 
Assume a codeword $\ket{\psi}  \in \mathcal{C}$ affected by a channel error described by the operator $\M{E} \in \mathcal{G}_n$. 
For error correction, the received state $\M{E}\ket{\psi}$ is measured according to the generators $\M{G}_1, \M{G}_2, \ldots, \M{G}_{n-k}$, resulting in a quantum error syndrome $\Syndrome(\M{E})=(s_1, s_2, \ldots,s_{n-k})$, with each $s_i =0$ or $1$ depending on the fact that $\M{E}$ commutes or anticommutes with $\M{G}_i$, respectively. Note that the syndrome depends on $\M{E}$ and not on the particular q-codeword $\ket{\psi}$. Moreover, measuring the syndrome does not change the quantum state, which remains $\M{E} \ket{\psi}$. 
Let $\mathcal{S}=\{\Syndrome^{(1)}, \Syndrome^{(2)}, \ldots, \Syndrome^{(m)}\}$ be the set of $m=2^{n-k}$ possible syndromes, with $\Syndrome^{(1)}=(0,0,\ldots,0)$ denoting the syndrome of the operators $\M{E}$ (including the identity $\M{I}$, i.e., the no-errors operator) such that $\M{E}\ket{\psi}$ is still a valid q-codeword.  
A generic Pauli error $\M{E} \in \mathcal{G}_n$ can be described by specifying the single Pauli errors on each qubit. We will use when necessary $\PauliXped{i},\PauliYped{i}$, and $\PauliZped{i}$ to denote the Pauli error $\PauliX, \PauliY$, and $\PauliZ$, respectively, on the $i$-th qubit of a codeword. 

\section{Hamming Bounds for Quantum Asymmetric Codes}
\label{sec:qhb}
The standard \ac{QHB} gives a necessary condition for the existence of non-degenerate codes able to correct generic errors. It states that  
a \ac{QECC} which encodes $k$ qubits in $n$ qubits can correct up to $t$ generic errors per codeword only if \cite{NieChu:2010,Eke:1996}
\begin{equation}
	\label{eq:QuantumHammingBound}
	2^{n-k} \ge \sum_{j = 0}^{t} \binom{n}{j}3^j \,.
\end{equation}
The bound is easily proved by noticing that the number of syndromes, $2^{n-k}$, must be at least equal to that of the distinct errors we want to correct. Since for each position there could be three Pauli errors ($\PauliX$, $\PauliY$ or $\PauliZ$), the number of distinct patterns having  
$j$ qubits in error is $\binom{n}{j} 3^j$, and this gives the bound \eqref{eq:QuantumHammingBound}. 

In this paper we investigate non-degenerate \acp{QECC} which can correct 
 some generic errors ($\PauliX$, $\PauliY$ or $\PauliZ$), plus some fixed Pauli type errors (e.g., $\PauliZ$ errors). 
We derive therefore the following \ac{GQHB}. 
\begin{theorem}[Generalized Quantum Hamming Bound] 
A quantum code which encodes $k$ qubits in $n$ qubits can correct up to $\eg$ generic errors plus up to $\eZ$ fixed Pauli type errors (e.g., $\PauliZ$ errors) per codeword only if
\begin{equation}
	\label{eq:QuantumHammingBound_eg}
	2^{n-k} \ge \sum_{j = 0}^{\eg+\eZ} \binom{n}{j}\sum_{i=0}^{\eg}  \binom{j}{i} 2^{i} \,.
\end{equation}
\end{theorem}
\begin{proof}
For the proof we need to enumerate the different patterns of error. The number of patterns of up to $\eg$ generic errors is given by \eqref{eq:QuantumHammingBound} with $t=\eg$.  
Then, we have to add the number of configurations with $\eg < j \le \eg+\eZ$ errors, composed by $\eg$ generic Pauli errors and the remaining $j-\eg$ Pauli $\PauliZ$ errors.  We can write %This number is  
\begin{equation}
	\label{eq:QuantumHammingBound_eg_proof1}
	2^{n-k} \ge \sum_{j = 0}^{\eg} \binom{n}{j}3^j + \sum_{j = \eg + 1}^{\eg+\eZ}\binom{n}{j}\left[ 3^j - f(j;\eg)\right]
\end{equation}
where $f(j;\eg)$ is a function that returns the number of non-correctable patterns of $j$ errors. This is the solution of the following combinatorial problem: given $j$ positions of the errors, 
count the number of all combinations with more than $\eg$ symbols from the set $\Set{P}_{\mathrm{XY}} = \left\{ \PauliX, \PauliY\right\}$ and the remaining from the set $\Set{P}_{\mathrm{Z}} = \left\{ \PauliZ \right\}$. We have therefore  
\begin{equation}
	\label{eq:QuantumHammingBound_eg_proof2}
	f(j;\eg) = \sum_{i = 0}^{j-\eg-1} \binom{j}{i} 2^{j-i}
\end{equation}
which allows to write 
\begin{align}
	\label{eq:QuantumHammingBound_eg_proof3}
	g(j;\eg) &= 3^j - f(j;\eg) \nonumber \\
	&= \sum_{i=0}^{j} \binom{j}{i} 2^{j-i} - \sum_{i = 0}^{j-\eg-1} \binom{j}{i} 2^{j-i} \nonumber \\
	&= \sum_{i=j-\eg}^{j} \binom{j}{i} 2^{j-i} = \sum_{i=0}^{\eg} \binom{j}{i} 2^{i}\,.
\end{align}
It is easy to see that $g(j;\eg)$ is equal to $3^j$ if $j \le \eg$, so substituting and incorporating the summation in \eqref{eq:QuantumHammingBound_eg_proof1} we finally obtain
\begin{equation}
	\label{eq:QuantumHammingBound_eg_proof4}
	 2^{n-k} \ge \sum_{j = 0}^{\eg+\eZ} \binom{n}{j}g(j;\eg) = \sum_{j = 0}^{\eg+\eZ} \binom{n}{j}\sum_{i=0}^{\eg} \binom{j}{i} 2^{i}\,.
\end{equation}
\end{proof}
The \ac{GQHB} in \eqref{eq:QuantumHammingBound_eg} can be used to compare codes which can correct $t$ generic errors with codes correcting a total of $t$ errors, with $\eg$ of them generic and the others $t-\eg$ of a fixed type. In Table~\ref{tab:ComparisonHammingBoundEG1} we report the minimum code lengths ${n}_{\mathrm{min}}$ resulting from the Hamming bounds, for different values of the total number of errors $t$, and assuming $\eg = 1$ for the \ac{GQHB}. From the table we can observe the possible gain in qubits for the asymmetric case. 
\begin{table}[tb]
	\centering
	\begin{tabular}{c c c c c c c c c}
		\toprule
		& \hspace{10pt} & $t = 1$ & \hspace{10pt} & $t = 2$ & \hspace{10pt} & $t = 3$ & \hspace{10pt} & $t = 4$ \\ 
		\midrule		
		$k = 1$ & & 5,5 & & 10,9 & & 15,12 & & 20,15 \\  
		$k = 2$ & & 7,7 & & 12,10 & & 16,14 & & 21,17\\  
		$k = 3$ & & 8,8 & & 13,12 & & 18,15 & & 23,19 \\  
		\bottomrule
	\end{tabular}
	\vspace{10pt}
	\caption{Comparison between the minimum code lengths $n_{\mathrm{min}}^{\textrm{QHB}}, n_{\mathrm{min}}^{\textrm{GQHB}}$ according to the Hamming bounds \eqref{eq:QuantumHammingBound} and \eqref{eq:QuantumHammingBound_eg}, respectively. For the \ac{GQHB} the bounds refer to $t=\eg+\eZ$ with $\eg = 1$.}
	\label{tab:ComparisonHammingBoundEG1}
\end{table}
%

%%%%%%%%%%%%%%%%%%%%%%%%%%
\section{Construction of short asymmetric codes by syndrome assignment}
In this section we present a construction of short stabilizer asymmetric codes with $k=1$ and $\eg = 1$, i.e.,  for $[[n,1]]$ \acp{QECC} with error correction capability $(1,\eZ)$. 
The design is based on the error syndromes: specifically, we proceed by assigning different syndromes to the different correctable error patters.  

Let us first observe that the vector syndrome of a composed error $\M{E} = \M{E}_1\M{E}_2$, with $\M{E}_1, \M{E}_2 \in \mathcal{G}_n$, can be expressed as $\Syndrome{\left(\M{E}\right)} = \Syndrome{\left(\M{E}_1\M{E}_2\right)} = \Syndrome{\left(\M{E}_2\M{E}_1\right)}=\Syndrome{\left(\M{E}_1\right)} \oplus \Syndrome{\left(\M{E}_2\right)}$ where $\oplus$ is the elementwise modulo $2$ addition. 
Moreover, $\PauliX \PauliZ=i \PauliY$, and for the syndromes we have 
 $\Syndrome{\left(\PauliXped{i} \PauliZped{i}\right)}=\Syndrome{\left(\PauliYped{i}\right)}$, $\Syndrome{\left(\PauliXped{i} \PauliYped{i}\right)}=\Syndrome{\left(\PauliZped{i}\right)}$, and  $\Syndrome{\left(\PauliYped{i} \PauliZped{i}\right)}=\Syndrome{\left(\PauliXped{i}\right)}$.
Hence, once we have assigned the syndromes for the single error patterns $\PauliXped{i}$ and $\PauliZped{i}$, with $i = 1,\dotsc, n$, the syndromes for all possible errors are automatically determined. 

In the following, if not specified otherwise, the indexes $i ,j $ will run from $1$ to $n$, and the index $\ell$ will run from $1$ to $n-1$. By definition the weight of a syndrome is the number of non-zero elements in the associated vector. 

\subsection{Construction of \normalfont{$[[n,1]]$} \acp{QECC} with $\eg = 1, \eZ=1$}
For this case we need to 
assign $2 n$ syndromes $\Syndrome{(\PauliXped{i})}$ and $\Syndrome{(\PauliZped{i})}$ 
 such that the syndromes of the errors $\PauliI$, $\PauliXped{i}$, $\PauliYped{i}$, $\PauliZped{i}$, $\PauliXped{i}\PauliZped{j}$, $\PauliYped{i}\PauliZped{j}$, and $\PauliZped{i}\PauliZped{j}$, are all different $ \forall i, j$ with $i \ne j$. 
We aim to construct the shortest possible code according to the \ac{GQHB}, i.e., a code with $n= 9$ (see Table \ref{tab:ComparisonHammingBoundEG1}). We start by assigning the syndromes of $\PauliZped{i}$ as reported in the Table~\ref{tab:FixZSyndromes}. 
In particular, we associate the all-one syndrome to a $\PauliZ$ error on the last qubit, i.e., $\Syndrome{\left(\PauliZped{9}\right)} = (1,1,...,1)$. Therefore, $\Syndrome{\left(\PauliZped{9}\right)}$ has weight $n-k=8$.
With this choice we have assigned all possible syndromes of weight $1$ and $8$. 
Also, the combinations of $\PauliZped{i}\PauliZped{j}$ with $i\ne j$, cover all possible syndromes of weight $2$ and $7$.\\
\begin{table}[tb]
	\centering
%	\resizebox{0.6\textwidth}{!}{%
	\begin{tabular}{c c c c c c c c c} 
		%\multicolumn{9}{c}{Syndromes assignment for $\PauliZped{i}$}\\
		\toprule
		& $s_8\,$ & $s_7\,$ & $s_6\,$ & $s_5\,$ & $s_4\,$ & $s_3\,$ & $s_2\,$ & $s_1\,$\\ 
		\midrule
		$\PauliZ_1$ & 0 & 0 & 0 & 0 & 0 & 0 & 0 & 1 \\
		$\PauliZ_2$ & 0 & 0 & 0 & 0 & 0 & 0 & 1 & 0 \\
		$\PauliZ_3$ & 0 & 0 & 0 & 0 & 0 & 1 & 0 & 0 \\
		$\PauliZ_4$ & 0 & 0 & 0 & 0 & 1 & 0 & 0 & 0 \\
		$\PauliZ_5$ & 0 & 0 & 0 & 1 & 0 & 0 & 0 & 0 \\
		$\PauliZ_6$ & 0 & 0 & 1 & 0 & 0 & 0 & 0 & 0 \\
		$\PauliZ_7$ & 0 & 1 & 0 & 0 & 0 & 0 & 0 & 0 \\
		$\PauliZ_8$ & 1 & 0 & 0 & 0 & 0 & 0 & 0 & 0 \\
		$\PauliZ_9$ & 1 & 1 & 1 & 1 & 1 & 1 & 1 & 1 \\
		\bottomrule
	\end{tabular}
%	}%
	\vspace{10pt}
	\caption{Assigned syndromes for single Pauli $\PauliZ$ errors.}
	\label{tab:FixZSyndromes}
\end{table}\\
To assign the syndromes of $\PauliXped{i}$ we then use a Monte Carlo approach. 
To reduce the search space, i.e., the set of possible syndromes, we observe the following: 
\begin{itemize}
\item The weight of $\Syndrome{\left(\PauliXped{i} \right)}$ cannot be $3$ or $6$. This is because otherwise $\Syndrome{\left(\PauliZped{j}\PauliXped{i}\right)}$ would have weight $2$ or $7$ for some $i$ and $j$, which are already assigned for errors of the type $\PauliZped{i}\PauliZped{j}$. Therefore the possible weights for $\Syndrome{\left(\PauliXped{i} \right)}$ are only $4$ and $5$. The same observation applies to $\Syndrome{\left(\PauliYped{i} \right)}$. We then fix the weight for $\Syndrome{\left(\PauliXped{i} \right)}$ equal to 4.
\item  
We can obtain $\Syndrome{\left(\PauliYped{\ell} \right)}$ with weight  $5$ for $\ell = 1,\dotsc, 8$, by imposing to ``0" the $\ell$-th element of the syndrome of $\PauliXped{\ell}$. Note that $\PauliY_9$ has  weight $4$ since $\PauliX_9$ has weight $4$. 
\end{itemize}
By following the previous rules, a possible assignment obtained by Monte Carlo is reported in Table~\ref{tab:XSyndromesEG1K1}. 

%\clearpage

\begin{table}[tb]
	\centering
%	\resizebox{0.6\textwidth}{!}{%
	\begin{tabular}{c c c c c c c c c}
	%	\multicolumn{9}{c}{{Syndromes Assignment for $\PauliXped{i}$}}\\
		\toprule
& $s_8\,$ & $s_7\,$ & $s_6\,$ & $s_5\,$ & $s_4\,$ & $s_3\,$ & $s_2\,$ & $s_1\,$\\ 
		\midrule
		$\PauliX_1$ & 1 & 0 & 1 & 1 & 1 & 0 & 0 & 0 \\
		$\PauliX_2$ & 1 & 0 & 0 & 1 & 0 & 1 & 0 & 1 \\
		$\PauliX_3$ & 0 & 0 & 1 & 0 & 1 & 0 & 1 & 1 \\
		$\PauliX_4$ & 1 & 1 & 1 & 0 & 0 & 1 & 0 & 0 \\
		$\PauliX_5$ & 0 & 1 & 0 & 0 & 1 & 1 & 0 & 1 \\
		$\PauliX_6$ & 1 & 1 & 0 & 0 & 0 & 0 & 1 & 1 \\
		$\PauliX_7$ & 0 & 0 & 1 & 1 & 0 & 1 & 1 & 0 \\
		$\PauliX_8$ & 0 & 1 & 0 & 1 & 1 & 0 & 1 & 0 \\
		$\PauliX_9$ & 1 & 0 & 0 & 0 & 1 & 1 & 1 & 0 \\
		\bottomrule
	\end{tabular}
%	}%
	\vspace{10pt}
	\caption{Syndromes for single Pauli $\PauliX$ errors.}
	\label{tab:XSyndromesEG1K1}
\end{table}
From Table~\ref{tab:FixZSyndromes} and Table~\ref{tab:XSyndromesEG1K1} we can then build the stabilizer matrix with the following procedure, where $s_{{j}}\left(\PauliXped{i}\right)$ indicates the $j$-th elements of the $\PauliXped{i}$'s syndrome:
\begin{itemize}
\item{if $s_{{j}}\left(\PauliXped{i}\right) = 0$ and $s_{{j}}\left(\PauliZped{i}\right) = 0$ put the element $\PauliI$ in position $(j,i)$ of the stabilizer matrix because it is the only Pauli operator which commutes  with both.}
\item{if $s_{{j}}\left(\PauliXped{i}\right) = 1$ and $s_{{j}}\left(\PauliZped{i}\right) = 0$ put the element $\PauliZ$ in position $(j,i)$ of the stabilizer matrix because it is the only Pauli operator which commutes  with $\PauliZ$ and anti-commute with $\PauliX$.}
\item{if $s_{{j}}\left(\PauliXped{i}\right) = 0$ and $s_{{j}}\left(\PauliZped{i}\right) = 1$ put the element $\PauliX$ in position $(j,i)$ of the stabilizer matrix because it is the only Pauli operator which commutes  with $\PauliX$ and anti-commute with $\PauliZ$.}
\item{if $s_{{j}}\left(\PauliXped{i}\right) = 1$ and $s_{{j}}\left(\PauliZped{i}\right) = 1$ put the element $\PauliY$ in position $(j,i)$ of the stabilizer matrix because it is the only Pauli operator which anti-commutes with both.}
\end{itemize}
The resulting stabilizer matrix, after checking the commutation conditions, is represented in Table \ref{tab:StabMatrixEG1K1}. According to the \ac{GQHB} the code specified in the table is therefore the shortest possible code for $k=1, \eg=1$ and $\eZ=1$. 

\begin{table}[tb]
	\centering
%	\resizebox{0.6\textwidth}{!}{%
	\begin{tabular}{c c c c c c c c c c c} 
		%\multicolumn{10}{c}{Stabilizers Matrix}\\
		\toprule
		&\hspace{15pt}& $1$ & $2$ & $3$ & $4$ & $5$ & $6$ & $7$ & $8$ & $9$\\
		\midrule
		$\M{G}_1$ & & $\PauliX$ & $\PauliZ$ & $\PauliZ$ & $\PauliI$ & $\PauliZ$ & $\PauliZ$ & $\PauliI$ & $\PauliI$ & $\PauliX$\\
		$\M{G}_2$ & & $\PauliI$ & $\PauliX$ & $\PauliZ$ & $\PauliI$ & $\PauliI$ & $\PauliZ$ & $\PauliZ$ & $\PauliZ$ & $\PauliY$\\
		$\M{G}_3$ & & $\PauliI$ & $\PauliZ$ & $\PauliX$ & $\PauliZ$ & $\PauliZ$ & $\PauliI$ & $\PauliZ$ & $\PauliI$ & $\PauliY$\\
		$\M{G}_4$ & & $\PauliZ$ & $\PauliI$ & $\PauliZ$ & $\PauliX$ & $\PauliZ$ & $\PauliI$ & $\PauliI$ & $\PauliZ$ & $\PauliY$\\
		$\M{G}_5$ & & $\PauliZ$ & $\PauliZ$ & $\PauliI$ & $\PauliI$ & $\PauliX$ & $\PauliI$ & $\PauliZ$ & $\PauliZ$ & $\PauliX$\\
		$\M{G}_6$ & & $\PauliZ$ & $\PauliI$ & $\PauliZ$ & $\PauliZ$ & $\PauliI$ & $\PauliX$ & $\PauliZ$ & $\PauliI$ & $\PauliX$\\
		$\M{G}_7$ & & $\PauliI$ & $\PauliI$ & $\PauliI$ & $\PauliZ$ & $\PauliZ$ & $\PauliZ$ & $\PauliX$ & $\PauliZ$ & $\PauliX$\\
		$\M{G}_8$ & & $\PauliZ$ & $\PauliZ$ & $\PauliI$ & $\PauliZ$ & $\PauliI$ & $\PauliZ$ & $\PauliI$ & $\PauliX$ & $\PauliY$\\
		\bottomrule
	\end{tabular}
%	}%
	\vspace{10pt}
	\caption{Stabilizer for a $[[9,1]]$ \ac{QECC} with $\eg = 1$ and $\eZ = 1$.}
	\label{tab:StabMatrixEG1K1}
\end{table}
\subsection{Construction of \normalfont{$[[n,1]]$} \acp{QECC} with $\eg = 1$ and $\eZ \geq 1$}
The construction presented in the previous section can be generalized to the case of more fixed errors, $\eZ \geq 1$. In this section we indicate $\tilde{t}=\eg+\eZ$. 
We start by adopting the same assignment proposed in Table~\ref{tab:FixZSyndromes} for a single $\PauliZ$ error, i.e., $\Syndrome{(\PauliZped{\ell})}$ has a $1$ in position $\ell$, $\ell \neq n$, and $\Syndrome{\left(\PauliZped{n}\right)} = (1,1,...,1)$. Note that $\Syndrome{\left(\PauliZped{n}\right)}$ has weight $n-k=n-1$.
In this way, it is easy to see that we use all possible syndromes with weight in the range $\big[0,\tilde{t}\big]$ and $\big[n-\tilde{t},n-1\big]$, covering all possible error operators with up to $\tilde{t}$ errors of type $\PauliZ$. %, $i = 1,\dotsc, n$.\\
For the assignment of the syndromes $\Syndrome{(\PauliXped{i})}$ we can generalize the previously exposed arguments, as follows:

\begin{itemize}
\item The weight of $\Syndrome{\left(\PauliXped{i} \right)}$ cannot be less than $2\tilde{t}$ or greater than $n-2\tilde{t}$. This is because otherwise $\Syndrome{\left(\PauliZped{j_1}\dotsc\PauliZped{j_{L}}\PauliXped{i}\right)}$ would have weight in the range $\big[0,\tilde{t}\big]$ or $\big[n-\tilde{t},n-1\big]$ for some $L\le\eZ$ and some choices of $j_1,\dotsc,j_{\mathrm{L}}$. These syndromes are already assigned for errors of the type $\PauliZped{j_1}\dotsc\PauliZped{j_M}$ for some $M\le\eZ$ and some choices of $j_1,\dotsc,j_{\mathrm{M}}$. Therefore the possible weights for $\Syndrome{\left(\PauliXped{i} \right)}$ are in the range $\big[2\tilde{t},n-2\tilde{t}\big]$. The same observation applies to $\Syndrome{\left(\PauliYped{i} \right)}$. %We then fix the weight for $\Syndrome{\left(\PauliXped{i} \right)}$ equal to 4.
\item
Setting the $\ell$-th element of the syndrome of $\PauliXped{\ell}$ to ``0" we obtain that $\Syndrome{\left(\PauliYped{\ell} \right)}$ has the weight of $\Syndrome{\left(\PauliXped{\ell} \right)}$ increased by 1, with $\ell = 1,\dotsc,n-1$. Hence, in order to have both $\Syndrome{\left(\PauliXped{\ell} \right)}$ and $\Syndrome{\left(\PauliYped{\ell} \right)}$ in the permitted range, we must have $n-4\tilde{t} \ge 1$. Note that this constraint can be stricter than the \ac{GQHB}. For example, we cannot construct the $[[12,1]]$ code with $\eg = 1, \eZ=2$. The comparision between the GQHB and the construnction bound $n-4\tilde{t} \ge 1$ is reported in Table \ref{tab:ComparisonHammingBoundConstruction}.
\item
About $\PauliXped{n}$ and $\PauliYped{n}$, we recall that $\Syndrome{\left(\PauliZped{n}\right)}=\Syndrome{\left(\PauliXped{n}\right)} \oplus  \Syndrome{\left(\PauliYped{n}\right)}$ and that we choose $\Syndrome{\left(\PauliZped{n}\right)} = (1,1,...,1)$. Therefore, in the positions where the syndrome of $\PauliXped{n}$ has a $1$, the syndrome of $\PauliYped{n}$ has a $0$, and viceversa. As a consequence, the sum of the weights of  $\Syndrome{\left(\PauliXped{n}\right)}$ and $\Syndrome{\left(\PauliYped{n}\right)}$ is $n-1$. 
Then, a good choice is to assign to  $\Syndrome{\left(\PauliXped{n} \right)}$ a weight $\lceil(n-1)/2\rceil$ or $\lfloor(n-1)/2\rfloor$. In this case, if $n$ is odd $\Syndrome{\left(\PauliYped{n} \right)}$ would have the same weight, which is in the correct range because $n-4\tilde{t} \ge 0$ is guaranteed by the previous point; if $n$ is even the weights are still in the correct range because $n-4\tilde{t} \ge 1$.
\end{itemize}

\begin{table}[htb]
	\centering
	%\resizebox{0.6\textwidth}{!}{%
	\begin{tabular}{c c c c c c c}
		%\multicolumn{5}{c}{}\\%{Comparison between ${n}_{\mathrm{min}}$}\\
		\toprule
		$\tilde t$ & $1$ & $2$ & $3$ & $4$ & $5$ & $6$\\ 
		\midrule		
		%\addlinespace
		GQHB & 5 & 9 & 12 & 15 & 18 & 21 \\   
		$1 + 4\tilde{t}$ & 5 & 9 & 13 & 17 & 21 & 25 \\  
		\bottomrule
	\end{tabular}
	\vspace{10pt}
	%}%
	\caption{Comparison between the minimum code lengths according to the generalized quantum Hamming bound  \eqref{eq:QuantumHammingBound_eg} and the construction bound $n >= 1+4\tilde t$. The bounds refer to $\tilde t=\eg+\eZ$ with $\eg = 1$.}
	\label{tab:ComparisonHammingBoundConstruction}
\end{table}
The procedure is summarized as Algorithm~\ref{alg:assignment}. For example, we obtained the $[[13,1]]$ \ac{QECC} with $\eg = 1$ and $\eZ = 2$ reported in Table~\ref{tab:StabMatrixEG1K1EZ2}. 

The proposed method to design asymmetric codes is based on a Monte Carlo search over a reduced syndrome assignments space. To give an idea of the time needed to find a valid code, we performed several runs of the algorithm. % to investigate the expected number of trials to obtain a valid construction.
 In our simulations the number of expected trials for the asymmetric $[[9,1]]$ case is around $20$, and for the asymmetric $[[13,1]]$ case is around $1100$.  

\begin{algorithm}[t]%[]
\SetAlgoHangIndent{0.5cm}
\KwResult{Stabilizer matrix.}

\medskip
Choose $n$ and $\tilde{t}$ to satisfy the constraint $n-4\tilde{t} \ge 1$\;
Assign $\Syndrome{\left(\PauliZped{i}\right)}$ as in Table \ref{tab:FixZSyndromes}\;
Pick a random syndrome for $\Syndrome{\left(\PauliXped{n} \right)}$ with weight $\lfloor(n-1)/2\rfloor$\;
Assign $\Syndrome{\left(\PauliYped{n}\right)}$, $\Syndrome{\left(\PauliXped{n}\PauliZped{j_1}\dotsc\PauliZped{j_{\mathrm{L}}} \right)}$ and $\Syndrome{\left(\PauliYped{n}\PauliZped{j_1}\dotsc\PauliZped{j_{\mathrm{L}}} \right)}$ for each $L=1,\dotsc,\eZ$ and for each possible combination of $j_1,\dotsc,j_{\mathrm{L}} \neq n$\;
\For{ $\ell = 1$ \KwTo $n-1$}{
	goodPick = 0\;
	\While{goodPick == 0}{
  		Pick a random syndrome for $\Syndrome{\left(\PauliXped{\ell} \right)}$ with weight in $\big[2\tilde{t},n-2\tilde{t}-1\big]$ and $s_{\ell}\left(\PauliXped{\ell}\right) = 0$\;
  		\If{$\Syndrome{\left(\PauliYped{\ell} \right)}$, $\Syndrome{\left(\PauliXped{\ell}\PauliZped{j_1}\dotsc\PauliZped{j_{\mathrm{L}}} \right)}$ and $\Syndrome{\left(\PauliYped{\ell}\PauliZped{j_1}\dotsc\PauliZped{j_{\mathrm{L}}} \right)}$ are not already assigned for all possible combinations}{
   			goodPick = 1\;
			Assign $\Syndrome{\left(\PauliYped{\ell} \right)}$ and all  $\Syndrome{\left(\PauliXped{\ell}\PauliZped{j_1}\dotsc\PauliZped{j_{\mathrm{L}}} \right)}$, $\Syndrome{\left(\PauliYped{\ell}\PauliZped{j_1}\dotsc\PauliZped{j_{\mathrm{L}}} \right)}$\;
   		}
		\If{no more possible syndromes}{
			Restart the algorithm\;
		}
	}
}
Construct the stabilizer matrix from $\Syndrome{\left(\PauliXped{i} \right)}$ and $\Syndrome{\left(\PauliZped{i} \right)}$\;
Check if all of the generators commute with each other. \\
\caption{Construction by syndrome assignment, $k=1, \eg=1$.}
\label{alg:assignment}	
\end{algorithm}

\begin{table*}[ht]
	\centering
	\begin{tabular}{c c c c c c c c c c c c c c c }
		\toprule
		&\hspace{15pt}& $1$ & $2$ & $3$ & $4$ & $5$ & $6$ & $7$ & $8$ & $9$ & $10$ & $11$ & $12$ & $13$ \\
		\midrule
		$\M{G}_{1}$ & & $\PauliX$ & $\PauliZ$ & $\PauliI$ & $\PauliZ$ & $\PauliZ$ & $\PauliZ$ & $\PauliI$ & $\PauliZ$ & $\PauliI$ & $\PauliI$ & $\PauliI$ & $\PauliZ$ & $\PauliX$  \\
		$\M{G}_{2}$ & & $\PauliI$ & $\PauliX$ & $\PauliZ$ & $\PauliI$ & $\PauliZ$ & $\PauliZ$ & $\PauliZ$ & $\PauliZ$ & $\PauliZ$ & $\PauliI$ & $\PauliI$ & $\PauliI$ & $\PauliY$  \\
		$\M{G}_{3}$ & & $\PauliZ$ & $\PauliZ$ & $\PauliX$ & $\PauliZ$ & $\PauliI$ & $\PauliI$ & $\PauliI$ & $\PauliZ$ & $\PauliZ$ & $\PauliZ$ & $\PauliI$ & $\PauliI$ & $\PauliY$  \\
		$\M{G}_{4}$ & & $\PauliI$ & $\PauliI$ & $\PauliZ$ & $\PauliX$ & $\PauliI$ & $\PauliZ$ & $\PauliI$ & $\PauliZ$ & $\PauliI$ & $\PauliZ$ & $\PauliZ$ & $\PauliZ$ & $\PauliY$  \\
		$\M{G}_{5}$ & & $\PauliZ$ & $\PauliI$ & $\PauliZ$ & $\PauliZ$ & $\PauliX$ & $\PauliI$ & $\PauliZ$ & $\PauliZ$ & $\PauliI$ & $\PauliI$ & $\PauliZ$ & $\PauliI$ & $\PauliX$  \\
		$\M{G}_{6}$ & & $\PauliI$ & $\PauliZ$ & $\PauliI$ & $\PauliZ$ & $\PauliZ$ & $\PauliX$ & $\PauliZ$ & $\PauliI$ & $\PauliI$ & $\PauliZ$ & $\PauliZ$ & $\PauliI$ & $\PauliY$  \\
		$\M{G}_{7}$ & & $\PauliI$ & $\PauliI$ & $\PauliZ$ & $\PauliZ$ & $\PauliZ$ & $\PauliI$ & $\PauliX$ & $\PauliI$ & $\PauliZ$ & $\PauliZ$ & $\PauliI$ & $\PauliZ$ & $\PauliX$  \\
		$\M{G}_{8}$ & & $\PauliZ$ & $\PauliI$ & $\PauliI$ & $\PauliI$ & $\PauliZ$ & $\PauliZ$ & $\PauliI$ & $\PauliX$ & $\PauliZ$ & $\PauliZ$ & $\PauliZ$ & $\PauliI$ & $\PauliX$  \\
		$\M{G}_{9}$ & & $\PauliZ$ & $\PauliZ$ & $\PauliZ$ & $\PauliI$ & $\PauliZ$ & $\PauliI$ & $\PauliI$ & $\PauliI$ & $\PauliX$ & $\PauliI$ & $\PauliZ$ & $\PauliZ$ & $\PauliY$  \\
		$\M{G}_{10}$ & & $\PauliI$ & $\PauliZ$ & $\PauliI$ & $\PauliI$ & $\PauliI$ & $\PauliI$ & $\PauliZ$ & $\PauliZ$ & $\PauliZ$ & $\PauliX$ & $\PauliZ$ & $\PauliZ$ & $\PauliX$  \\
		$\M{G}_{11}$ & & $\PauliZ$ & $\PauliI$ & $\PauliI$ & $\PauliZ$ & $\PauliI$ & $\PauliZ$ & $\PauliZ$ & $\PauliI$ & $\PauliZ$ & $\PauliI$ & $\PauliX$ & $\PauliZ$ & $\PauliY$  \\
		$\M{G}_{12}$ & & $\PauliZ$ & $\PauliZ$ & $\PauliZ$ & $\PauliI$ & $\PauliI$ & $\PauliZ$ & $\PauliZ$ & $\PauliI$ & $\PauliI$ & $\PauliZ$ & $\PauliI$ & $\PauliX$ & $\PauliX$  \\
		\bottomrule
	\end{tabular}
	\vspace{10pt}
	\caption{Stabilizer for a $[[13,1]]$ \ac{QECC} with $\eg = 1$ and $\eZ = 2$.}
	\label{tab:StabMatrixEG1K1EZ2}
\end{table*}			
\section{Performance Analysis over Asymmetric Channels}

It is well known that the \ac{CWEP} for a standard $[[n,k]]$ \ac{QECC} which corrects up to $t$ generic errors per codeword is
\begin{equation}
\label{eq:PeGen}
P_{\mathrm e} = 1 - \sum_{j = 0}^{t} \binom{n}{j}(1-\rho)^{n-j}\rho^j 
\end{equation}
where $\rho=\px+\py+\pz$ is the qubit error probability. 

We now generalize this expression to an $[[n,k]]$ \ac{QECC} which corrects up to $\eg$ generic errors and up to $\eZ$ Pauli $\PauliZ$ errors per codeword, over a generic asymmetric channel with Pauli error probabilities $\px, \py$ and $\pz$. 
To this aim, we first note that the patterns of correctable errors are those discussed in Section~\ref{sec:qhb}. Then, by weighting each pattern with the corresponding probability of occurrence, it is not difficult to show that the performance in terms of \ac{CWEP} is %then given by the following equation
\begin{equation}
\label{eq:PeGen}
P_{\mathrm e} = 1 - \sum_{j = 0}^{\eg+\eZ} \binom{n}{j}(1-\rho)^{n-j}\xi(j;\eg)
\end{equation}
where
\begin{align}
\label{eq:PeGenxi}
\xi&(j;\eg) = \nonumber\\
&\begin{dcases}
\rho^j & \text{if}~j\le\eg\\
\sum_{i = j-\eg}^{j}\binom{j}{i} \, \pz^i \, \sum_{\ell = 0}^{j-i} \binom{j-i}{\ell}\, \px^\ell \, \py^{j-i-\ell} & \text{otherwise}\,.\\
\end{dcases}
\end{align}
In the case of asymmetric channels with $\px=\py=\rho/(A+2)$, $\pz = A\rho/(A+2)$, and $A=\pz/\px$, the expression in \eqref{eq:PeGen} can be simplified to
\begin{align}
\label{eq:PePart}
P_{\mathrm e} =&\: 1 - \sum_{j = 0}^{\eg+\eZ} \binom{n}{j}(1-\rho)^{n-j}\rho^j \nonumber \\
&\times\left( 1 -  2^{j+1}\frac{(A/2)^{j-\eg} - 1}{(A-2)(A+2)^{j}} {u_{j-\eg - 1}}\right)
\end{align}
where $u_{i} = 1$ if $i\ge0$, otherwise $u_{i} = 0$.

\begin{figure}[ht]
	\centering
	\resizebox{0.98\columnwidth}{!}{%
\begin{tikzpicture}

\begin{axis}[%
width=4.438in,
height=3.365in,
at={(0.841in,0.682in)},
scale only axis,
xmode=log,
xmin=0.001,
xmax=0.05,
xtick={0.001, 0.002, 0.005,  0.01,  0.02,  0.05,   0.1},
xticklabels={$10^{-3}$, $2\cdot10^{-3}$, $5\cdot10^{-3}$,  $10^{-2}$,  $2\cdot10^{-2}$,  $5\cdot10^{-2}$,   0.1},
xminorticks=true,
xlabel style={font=\color{white!15!black}},
xlabel={$\rho$},
ymode=log,
ymin=1e-05,
ymax=0.1,
ytick={ 1e-05, 0.0001,  0.001,   0.01,    0.1},
yminorticks=true,
ylabel style={font=\color{white!15!black}},
ylabel={CWER},
axis background/.style={fill=white},
legend style={at={(0.03,0.97)}, anchor=north west, legend cell align=left, align=left, draw=white!15!black}
]
\addplot [color=black, dashed, line width=1.0pt]
  table[row sep=crcr]{%
0.0001	9.99800014432096e-08\\
0.0002	3.99840023889743e-07\\
0.0003	8.99460121351345e-07\\
0.0004	1.59872038372642e-06\\
0.0005	2.49750093712703e-06\\
0.0006	3.59568194387497e-06\\
0.0007	4.89314360104399e-06\\
0.0008	6.38976614279925e-06\\
0.0009	8.08542983915701e-06\\
0.001	9.98001499598767e-06\\
0.002	3.98402398720053e-05\\
0.003	8.94612140279975e-05\\
0.004	0.00015872383590404\\
0.005	0.000247509362500023\\
0.006	0.000355699408895998\\
0.007	0.000483175947772072\\
0.008	0.000629821308928011\\
0.009	0.000795518178803988\\
0.01	0.000980149600000091\\
0.015	0.00218325633750009\\
0.02	0.00384238720000007\\
0.03	0.0084720528000001\\
0.04	0.0147579904000001\\
0.05	0.0225925000000003\\
0.06	0.0318712896000003\\
0.07	0.0424934272000003\\
0.08	0.0543612927999998\\
0.09	0.0673805303999997\\
0.1	0.0814599999999999\\
};
\addlegendentry{Code [[5,1]]}

\addplot [color=black, dashdotted, line width=1.0pt]
  table[row sep=crcr]{%
0.0001	1.64900953624271e-10\\
0.0002	1.31841663092158e-09\\
0.0003	4.44698741472321e-09\\
0.0004	1.0534683858832e-08\\
0.0005	2.05632108957403e-08\\
0.0006	3.55119118264717e-08\\
0.0007	5.63577667188662e-08\\
0.0008	8.40754034246824e-08\\
0.0009	1.19637095527779e-07\\
0.001	1.64012767430133e-07\\
0.002	1.30424840901239e-06\\
0.003	4.37548023883452e-06\\
0.004	1.03093796854395e-05\\
0.005	2.00148406979499e-05\\
0.006	3.43783009012469e-05\\
0.007	5.42640595223051e-05\\
0.008	8.05145921138118e-05\\
0.009	0.000113950862097807\\
0.01	0.000155372629155111\\
0.015	0.000508809449745158\\
0.02	0.00117018096808067\\
0.03	0.00371719763319261\\
0.04	0.0082913179793446\\
0.05	0.0152352973052983\\
0.06	0.024762846606895\\
0.07	0.0369792909298049\\
0.08	0.0518999825205937\\
0.09	0.0694666496177746\\
0.1	0.0895618508499997\\
};
\addlegendentry{Code [[11,1]]}

\addplot [color=black, dotted, line width=1.0pt]
  table[row sep=crcr]{%
0.0001	3.59832037745466e-07\\
0.0002	1.43865660448615e-06\\
0.0003	3.23546706032605e-06\\
0.0004	5.74925767125563e-06\\
0.0005	8.97902360872879e-06\\
0.0006	1.29237609500258e-05\\
0.0007	1.75824666734763e-05\\
0.0008	2.29541386639609e-05\\
0.0009	2.90377757085781e-05\\
0.001	3.58323774964538e-05\\
0.002	0.000142662031898819\\
0.003	0.000319494495833773\\
0.004	0.000565344253620849\\
0.005	0.000879234681545736\\
0.006	0.00126019798843117\\
0.007	0.00170727515650709\\
0.008	0.00221951588257656\\
0.009	0.00279597851947888\\
0.01	0.00343573001784638\\
0.015	0.00755175827231815\\
0.02	0.0131148938051288\\
0.03	0.028158234297257\\
0.04	0.04776575574134\\
0.05	0.0712113961953129\\
0.06	0.0978379709591997\\
0.07	0.127052376922538\\
0.08	0.158321048054399\\
0.09	0.191165652501938\\
0.1	0.225159022\\
};
\addlegendentry{$\mathcal{C}_{\mathrm S}$ \hspace{1.57pt} [[9,1]]}

\addplot [color=black, line width=1.0pt, mark=triangle, mark options={solid, rotate=180, black}]
  table[row sep=crcr]{%
0.0001	2.49909014649315e-07\\
0.0002	9.99272235009325e-07\\
0.0003	2.24754419016132e-06\\
0.0004	3.99417976159645e-06\\
0.0005	6.23863418325403e-06\\
0.0006	8.98036304234548e-06\\
0.0007	1.22188222750379e-05\\
0.0008	1.59534681724168e-05\\
0.0009	2.01837573766122e-05\\
0.001	2.4909146881069e-05\\
0.002	9.92743481942082e-05\\
0.003	0.000222554878108613\\
0.004	0.000394213510287669\\
0.005	0.000613716503673128\\
0.006	0.000880533588293144\\
0.007	0.00119413795109702\\
0.008	0.00155400622181427\\
0.009	0.00195961845884065\\
0.01	0.00241045813514869\\
0.015	0.00532522691057412\\
0.02	0.00929514145884327\\
0.03	0.0201592041339238\\
0.04	0.0345402801266363\\
0.05	0.0520071205527348\\
0.06	0.0721582582291711\\
0.07	0.0946206999967899\\
0.08	0.11904864732735\\
0.09	0.145122245677202\\
0.1	0.172546363\\
};
\addlegendentry{$\mathcal{C}_{\mathrm{SE}}$ [[9,1]] A = 1}

\addplot [color=black, line width=1.0pt, mark=o, mark options={solid, black}]
  table[row sep=crcr]{%
0.0001	2.19530288266379e-07\\
0.0002	8.77842372899367e-07\\
0.0003	1.97451823331579e-06\\
0.0004	3.50914004838155e-06\\
0.0005	5.48129019657737e-06\\
0.0006	7.89055125695018e-06\\
0.0007	1.07365060049241e-05\\
0.0008	1.40187374183901e-05\\
0.0009	1.77368286739599e-05\\
0.001	2.18903631473627e-05\\
0.002	8.72835701522068e-05\\
0.003	0.000195764292846387\\
0.004	0.000346919195766499\\
0.005	0.000540336934515607\\
0.006	0.000775608154073181\\
0.007	0.00105232548705642\\
0.008	0.00137008355193087\\
0.009	0.00172847895117336\\
0.01	0.00212711026938496\\
0.015	0.00470989460696486\\
0.02	0.00823950081041526\\
0.03	0.0179485630706026\\
0.04	0.0308852395931183\\
0.05	0.0466997571024223\\
0.06	0.0650613194383269\\
0.07	0.0856577638282196\\
0.08	0.108195183853694\\
0.09	0.132397522336547\\
0.1	0.15800613724\\
};
\addlegendentry{$\mathcal{C}_{\mathrm{SE}}$ [[9,1]] A = 3}

\addplot [color=black, line width=1.0pt, mark=diamond, mark options={solid, black}]
  table[row sep=crcr]{%
0.0001	2.3491951149984e-07\\
0.0002	9.39356184626122e-07\\
0.0003	2.11282743513885e-06\\
0.0004	3.75485095573384e-06\\
0.0005	5.86494471614383e-06\\
0.0006	8.44262696402543e-06\\
0.0007	1.14874162207054e-05\\
0.0008	1.49988312872063e-05\\
0.0009	1.89763912404351e-05\\
0.001	2.34196154335165e-05\\
0.002	9.33578458708522e-05\\
0.003	0.000209335839327909\\
0.004	0.000370877499833144\\
0.005	0.000577509479417772\\
0.006	0.000828761170092504\\
0.007	0.00112416469581383\\
0.008	0.00146325490443759\\
0.009	0.00184556935966271\\
0.01	0.00227064833296264\\
0.015	0.00502160899760902\\
0.02	0.00877426613889524\\
0.03	0.0190684272934693\\
0.04	0.0327368061791768\\
0.05	0.0493883556923832\\
0.06	0.0686564792205309\\
0.07	0.0901981985978243\\
0.08	0.11369331995548\\
0.09	0.13884359929201\\
0.1	0.1653719095\\
};
\addlegendentry{$\mathcal{C}_{\mathrm{SE}}$ [[9,1]] A = 10}

\addplot [color=blue, line width=1.0pt, mark=triangle, mark options={solid, rotate=180, blue}]
  table[row sep=crcr]{%
0.0001	1.59971995752465e-07\\
0.0002	6.39775932710105e-07\\
0.0003	1.43924366002654e-06\\
0.0004	2.55820692642076e-06\\
0.0005	3.99649738059286e-06\\
0.0006	5.75394657242519e-06\\
0.0007	7.83038594904295e-06\\
0.0008	1.02256468611535e-05\\
0.0009	1.29395605595493e-05\\
0.001	1.59719581957542e-05\\
0.002	6.37753342540724e-05\\
0.003	0.000143240645424391\\
0.004	0.000254197447560521\\
0.005	0.000396474358140994\\
0.006	0.000569899079089306\\
0.007	0.000774298419397875\\
0.008	0.00100949831755422\\
0.009	0.00127532386377301\\
0.01	0.0015715993220324\\
0.015	0.00350351943278355\\
0.02	0.00616988953915508\\
0.03	0.0136145430911967\\
0.04	0.0237194364418788\\
0.05	0.0362945313906254\\
0.06	0.0511475841773296\\
0.07	0.0680856916029961\\
0.08	0.0869166830961292\\
0.09	0.107450367366054\\
0.1	0.129499642\\
};
\addlegendentry{$\mathcal{C}_{\mathrm{A}}$ \ [[9,1]] A = 1}

\addplot [color=green, line width=1.0pt, mark=o, mark options={solid, green}]
  table[row sep=crcr]{%
0.0001	5.76436542520485e-08\\
0.0002	2.3074902876077e-07\\
0.0003	5.19577279073188e-07\\
0.0004	9.24388945065311e-07\\
0.0005	1.44544395178726e-06\\
0.0006	2.08300161109366e-06\\
0.0007	2.83732061813309e-06\\
0.0008	3.7086590581161e-06\\
0.0009	4.69727440324653e-06\\
0.001	5.803423513796e-06\\
0.002	2.33853450599623e-05\\
0.003	5.29986740147872e-05\\
0.004	9.48902828576331e-05\\
0.005	0.000149301072557766\\
0.006	0.000216466037506273\\
0.007	0.000296614329997955\\
0.008	0.000389969324262777\\
0.009	0.000496748680051602\\
0.01	0.000617164405775637\\
0.015	0.00143082114694184\\
0.02	0.00261404735497652\\
0.03	0.00616817319369378\\
0.04	0.0114077209605547\\
0.05	0.0184170966106254\\
0.06	0.0272421061450121\\
0.07	0.0378947487193907\\
0.08	0.0503576482374952\\
0.09	0.0645881413764811\\
0.1	0.0805220394399997\\
};
\addlegendentry{$\mathcal{C}_{\mathrm{A}}$ \ [[9,1]] A = 3}

\addplot [color=red, line width=1.0pt, mark=diamond, mark options={solid, red}]
  table[row sep=crcr]{%
0.0001	1.00769642577144e-08\\
0.0002	4.06154288780716e-08\\
0.0003	9.20761098019043e-08\\
0.0004	1.64918867794613e-07\\
0.0005	2.59602709490904e-07\\
0.0006	3.76585789224707e-07\\
0.0007	5.16325405717962e-07\\
0.0008	6.79278009047933e-07\\
0.0009	8.65899197777679e-07\\
0.001	1.0766437202295e-06\\
0.002	4.61031102051267e-06\\
0.003	1.10502576173542e-05\\
0.004	2.08373430152752e-05\\
0.005	3.44041155874373e-05\\
0.006	5.21748970829098e-05\\
0.007	7.45658665659615e-05\\
0.008	0.00010198514378746\\
0.009	0.000134832871993605\\
0.01	0.00017350130017191\\
0.015	0.0004673403031326\\
0.02	0.000961136339674762\\
0.03	0.00270677468665141\\
0.04	0.00568469696728301\\
0.05	0.0101068827871098\\
0.06	0.0161297940909271\\
0.07	0.0238606776133397\\
0.08	0.0333634093774271\\
0.09	0.0446639035141407\\
0.1	0.0577551069999997\\
};
\addlegendentry{$\mathcal{C}_{\mathrm{A}}$ \ [[9,1]] A = 10}

\end{axis}
\end{tikzpicture}%
	}%
	\caption{Performance of short codes over an asymmetric channel, $k=1$. Code $[[5,1]]$:  $5$-qubits code with $t=1$. Code $[[11,1]]$: $11$-qubits code with $t=2$. $\mathcal{C}_{\mathrm S}$~$[[9,1]]$: $9$-qubits code from [15] used with $t=1$. $\mathcal{C}_{\mathrm {SE}}$~$[[9,1]]$: $9$-qubits code from [15] used also for the correctable two-errors patterns with at least one $\PauliZ$. $\mathcal{C}_{\mathrm A}$~$[[9,1]]$: $9$-qubits asymmetric code with $\eg = 1, \eZ=1$.}
	\label{fig:PerformanceAsymCodes}
\end{figure}
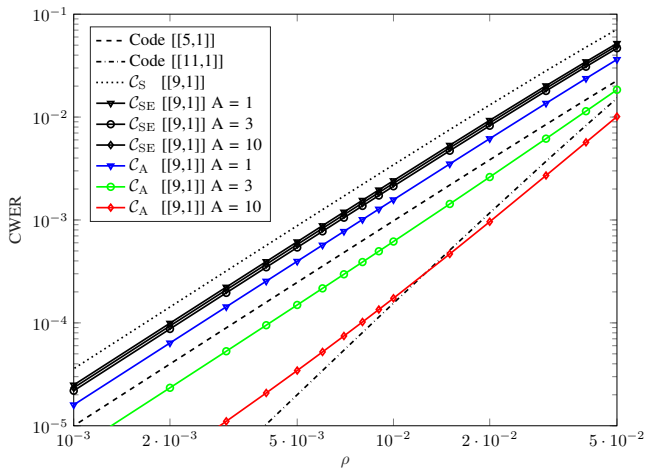

Using the previous expressions, we report in Fig.~\ref{fig:PerformanceAsymCodes} the performance in terms of \ac{CWEP} for different codes, assuming an asymmetric channel.  
The parameter $A$ accounts for the asymmetry of the channel, and for $A = 1$ we have the standard depolarizing channel.
In the figure we plot the \ac{CWEP} for the new asymmetric $[[9,1]]$ code specified in Table~\ref{tab:StabMatrixEG1K1} with $\eg = 1$ and $\eZ = 1$, reported in the plot as $\mathcal{C}_{\mathrm A}$, over channels with asymmetry parameter $A=1, 3$ and $10$. 
For comparison, in the same figure we report the \ac{CWEP} for the known $5$-qubits code, the Shor's $9$-qubits code (indicated in the figure as {$\mathcal{C}_{\mathrm S}$}), both correcting $t=1$ generic errors, and a $[[11,1]]$ code with $t=2$ \cite{Gra:07}. 
Additionally, for a fair comparison, we analyzed the Shor's code when used in an extended mode for patterns of two errors.  
In fact, besides arbitrary single errors, the Shor's code can also correct some combinations of multiple errors. 
The most relevant multiple errors for our asymmetric channel are in the form $\PauliZped{i}\PauliZped{j}$, $\PauliXped{i}\PauliZped{j}$, and $\PauliYped{i}\PauliZped{j}$. In this regard, it can be verified that the Shor code can be used to correct $9$ out of the $36$ possible $\PauliZped{i}\PauliZped{j}$ errors, all $72$ possible $\PauliXped{i}\PauliZped{j}$ errors, and $18$ out of the $72$ possible $\PauliYped{i}\PauliZped{j}$ errors. 
The Shor's code used with this extended error correction capability is reported in the following as $\mathcal{C}_{\mathrm{ SE}}$. 

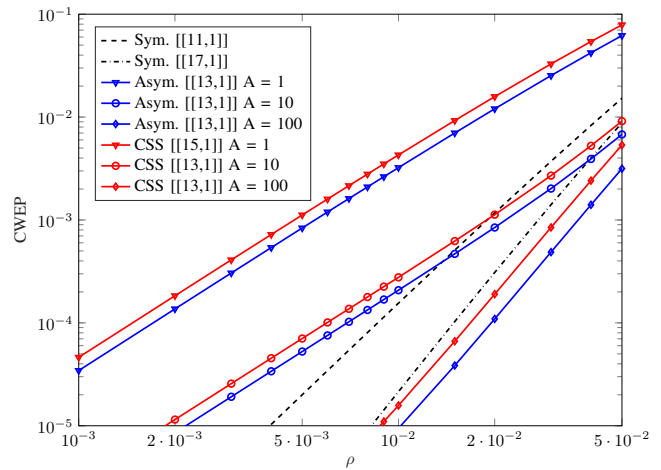
\begin{figure}[ht]
	\centering
	\resizebox{0.98\columnwidth}{!}{%
\begin{tikzpicture}

\begin{axis}[%
width=4.438in,
height=3.365in,
at={(0.841in,0.682in)},
scale only axis,
xmode=log,
xmin=0.001,
xmax=0.05,
xtick={0.001, 0.002, 0.005,  0.01,  0.02,  0.05,   0.1},
xticklabels={$10^{-3}$, $2\cdot10^{-3}$, $5\cdot10^{-3}$,  $10^{-2}$,  $2\cdot10^{-2}$,  $5\cdot10^{-2}$,   0.1},
xminorticks=true,
xlabel style={font=\color{white!15!black}},
xlabel={$\rho$},
ymode=log,
ymin=1e-05,
ymax=0.1,
ytick={ 1e-05, 0.0001,  0.001,   0.01,    0.1},
yminorticks=true,
ylabel style={font=\color{white!15!black}},
ylabel={CWEP},
axis background/.style={fill=white},
legend style={at={(0.03,0.97)}, anchor=north west, legend cell align=left, align=left, draw=white!15!black}
]
\addplot [color=black, dashed, line width=1.0pt]
  table[row sep=crcr]{%
0.0001	1.64900981758365e-10\\
0.0002	1.31841659811727e-09\\
0.0003	4.44698744495042e-09\\
0.0004	1.05346839029963e-08\\
0.0005	2.0563210911817e-08\\
0.0006	3.55119118555791e-08\\
0.0007	5.63577666756743e-08\\
0.0008	8.40754034170388e-08\\
0.0009	1.19637095563618e-07\\
0.001	1.64012767478461e-07\\
0.002	1.30424840905086e-06\\
0.003	4.37548023890511e-06\\
0.004	1.03093796854914e-05\\
0.005	2.00148406979128e-05\\
0.006	3.43783009012499e-05\\
0.007	5.42640595223576e-05\\
0.008	8.05145921138894e-05\\
0.009	0.000113950862097756\\
0.01	0.000155372629155104\\
0.015	0.000508809449745184\\
0.02	0.00117018096808075\\
0.03	0.00371719763319267\\
0.04	0.00829131797934457\\
0.05	0.0152352973052983\\
0.06	0.024762846606895\\
0.07	0.0369792909298048\\
0.08	0.0518999825205938\\
0.09	0.0694666496177746\\
0.1	0.0895618508499997\\
};
\addlegendentry{Sym. [[11,1]]}

\addplot [color=black, dashdotted, line width=1.0pt]
  table[row sep=crcr]{%
0.0001	2.37587727269783e-13\\
0.0002	3.7997383017796e-12\\
0.0003	1.92172944224467e-11\\
0.0004	6.06742434072771e-11\\
0.0005	1.47977519127096e-10\\
0.0006	3.06529912563747e-10\\
0.0007	5.6729310138337e-10\\
0.0008	9.66769997390315e-10\\
0.0009	1.54696799814502e-09\\
0.001	2.35537145343301e-09\\
0.002	3.72958071093166e-08\\
0.003	1.86854640094047e-07\\
0.004	5.8443455563939e-07\\
0.005	1.41205369241781e-06\\
0.006	2.89767512628991e-06\\
0.007	5.31261790159832e-06\\
0.008	8.96904779124075e-06\\
0.009	1.42175460112748e-05\\
0.01	2.14447541400098e-05\\
0.015	0.000103036825517266\\
0.02	0.00030903747237232\\
0.03	0.00140857327070576\\
0.04	0.00400662473555702\\
0.05	0.00880060494768631\\
0.06	0.0164130791202717\\
0.07	0.0273399978810487\\
0.08	0.041924408848889\\
0.09	0.0603488048906541\\
0.1	0.0826406225560634\\
};
\addlegendentry{Sym. [[17,1]]}

\addplot [color=blue, line width=1.0pt, mark=triangle, mark options={solid, rotate=180, blue}]
  table[row sep=crcr]{%
0.0001	3.46412579399491e-07\\
0.0002	1.38463504795183e-06\\
0.0003	3.11314692647002e-06\\
0.0004	5.53043096551686e-06\\
0.0005	8.63497313718931e-06\\
0.0006	1.2425262632898e-05\\
0.0007	1.68997918496006e-05\\
0.0008	2.20570563933542e-05\\
0.0009	2.78955550668814e-05\\
0.001	3.44137898652397e-05\\
0.002	0.000136654333892983\\
0.003	0.000305244153714246\\
0.004	0.000538736969397324\\
0.005	0.000835717105317046\\
0.006	0.00119479890528407\\
0.007	0.00161462615681507\\
0.008	0.00209387152443063\\
0.009	0.00263123599187187\\
0.01	0.00322544831311844\\
0.015	0.00700661067325736\\
0.02	0.0120333288324302\\
0.03	0.0253140848213903\\
0.04	0.0421776896952457\\
0.05	0.0619122688666248\\
0.06	0.0839492674728718\\
0.07	0.107834328456111\\
0.08	0.133203130657854\\
0.09	0.159761514929977\\
0.1	0.1872692981758\\
};
\addlegendentry{Asym. [[13,1]] A = 1}

\addplot [color=blue, line width=1.0pt, mark=o, mark options={solid, blue}]
  table[row sep=crcr]{%
0.0001	2.16508531103443e-08\\
0.0002	8.65407613348523e-08\\
0.0003	1.94577100320892e-07\\
0.0004	3.45669045498376e-07\\
0.0005	5.39727564308201e-07\\
0.0006	7.76665413093625e-07\\
0.0007	1.05639712377759e-06\\
0.0008	1.37883900763747e-06\\
0.0009	1.74390914253753e-06\\
0.001	2.15152736882107e-06\\
0.002	8.55146745371016e-06\\
0.003	1.913089379868e-05\\
0.004	3.3837783877444e-05\\
0.005	5.26364312335437e-05\\
0.006	7.55068693022443e-05\\
0.007	0.000102444306752436\\
0.008	0.000133458574186229\\
0.009	0.000168573582036813\\
0.01	0.000207826789505705\\
0.015	0.000468360659876765\\
0.02	0.000844878530116722\\
0.03	0.00201885665277957\\
0.04	0.00391878760272157\\
0.05	0.00677857571175744\\
0.06	0.010848424187912\\
0.07	0.0163729653197258\\
0.08	0.023575068503922\\
0.09	0.0326444114926826\\
0.1	0.0437300070012996\\
};
\addlegendentry{Asym. [[13,1]] A = 10}

\addplot [color=blue, line width=1.0pt, mark=diamond, mark options={solid, blue}]
  table[row sep=crcr]{%
0.0001	2.99736013786855e-10\\
0.0002	1.19892229477614e-09\\
0.0003	2.69880728875904e-09\\
0.0004	4.80234485511488e-09\\
0.0005	7.51418760547296e-09\\
0.0006	1.08406823517981e-08\\
0.0007	1.47898581159822e-08\\
0.0008	1.93714285723345e-08\\
0.0009	2.45967807233072e-08\\
0.001	3.04789693483798e-08\\
0.002	1.29479694854773e-07\\
0.003	3.20679478971186e-07\\
0.004	6.43719777593255e-07\\
0.005	1.15361875052677e-06\\
0.006	1.92019565636858e-06\\
0.007	3.02750692138254e-06\\
0.008	4.57329372038906e-06\\
0.009	6.66844090579932e-06\\
0.01	9.43644712292357e-06\\
0.015	3.85103129761921e-05\\
0.02	0.000109305672871152\\
0.03	0.000487336461763688\\
0.04	0.00140349646169047\\
0.05	0.00315386924136829\\
0.06	0.00604248639409111\\
0.07	0.010359934525292\\
0.08	0.0163676872550477\\
0.09	0.0242872316819263\\
0.1	0.0342931677891353\\
};
\addlegendentry{Asym. [[13,1]] A = 100}

\addplot [color=red, line width=1.0pt, mark=triangle, mark options={solid, rotate=180, red}]
  table[row sep=crcr]{%
0.0001	4.66262473003582e-07\\
0.0002	1.86343512461517e-06\\
0.0003	4.18910031066689e-06\\
0.0004	7.44084638594911e-06\\
0.0005	1.16162676915488e-05\\
0.0006	1.67129645419634e-05\\
0.0007	2.27285432059943e-05\\
0.0008	2.96606159028485e-05\\
0.0009	3.75068007822494e-05\\
0.001	4.62647219128729e-05\\
0.002	0.000183470927513461\\
0.003	0.000409280671770735\\
0.004	0.000721413621248726\\
0.005	0.00111764576375287\\
0.006	0.00159580814218908\\
0.007	0.00215378561225414\\
0.008	0.00278951562359622\\
0.009	0.00350098702407595\\
0.01	0.00428623888677119\\
0.015	0.00925394531855213\\
0.02	0.0157989515299359\\
0.03	0.0328640342018424\\
0.04	0.0541884197029595\\
0.05	0.0787766372165672\\
0.06	0.105863108640387\\
0.07	0.13485770992055\\
0.08	0.165302448015956\\
0.09	0.196837391439877\\
0.1	0.229174247947996\\
};
\addlegendentry{CSS [[15,1]] A = 1}

\addplot [color=red, line width=1.0pt, mark=o, mark options={solid, red}]
  table[row sep=crcr]{%
0.0001	2.91415063293804e-08\\
0.0002	1.1646632532173e-07\\
0.0003	2.61827015361264e-07\\
0.0004	4.65078939628575e-07\\
0.0005	7.26080254562324e-07\\
0.0006	1.04469190053101e-06\\
0.0007	1.42077758472286e-06\\
0.0008	1.85420378045664e-06\\
0.0009	2.34483970894075e-06\\
0.001	2.892557331347e-06\\
0.002	1.14829953372221e-05\\
0.003	2.56606895291919e-05\\
0.004	4.53411419447467e-05\\
0.005	7.04649486740619e-05\\
0.006	0.000100996791270791\\
0.007	0.000136924451644144\\
0.008	0.000178257850043048\\
0.009	0.000225028105741558\\
0.01	0.000277286620043892\\
0.015	0.000624010879858623\\
0.02	0.0011258075408812\\
0.03	0.0026984820204588\\
0.04	0.00526002817231138\\
0.05	0.0091280040424668\\
0.06	0.0146287286221086\\
0.07	0.0220654744156381\\
0.08	0.0316966744942237\\
0.09	0.0437222006194841\\
0.1	0.0582760482537324\\
};
\addlegendentry{CSS [[13,1]] A = 10}

\addplot [color=red, line width=1.0pt, mark=diamond, mark options={solid, red}]
  table[row sep=crcr]{%
0.0001	4.0347221295909e-10\\
0.0002	1.61406562031699e-09\\
0.0003	3.63439977734749e-09\\
0.0004	6.47022468387117e-09\\
0.0005	1.01304064570986e-08\\
0.0006	1.46269157780463e-08\\
0.0007	1.99748075652099e-08\\
0.0008	2.6192216966539e-08\\
0.0009	3.33003400283623e-08\\
0.001	4.13234210236821e-08\\
0.002	1.79363535096053e-07\\
0.003	4.57689761575393e-07\\
0.004	9.48769122018624e-07\\
0.005	1.75267446713899e-06\\
0.006	2.99582495610628e-06\\
0.007	4.82975820622774e-06\\
0.008	7.42993354994983e-06\\
0.009	1.09945658340475e-05\\
0.01	1.5743489204461e-05\\
0.015	6.63229743072518e-05\\
0.02	0.000190059289658816\\
0.03	0.000846076490506558\\
0.04	0.00241213013657897\\
0.05	0.00535156117184184\\
0.06	0.0101118919400456\\
0.07	0.0170897196119895\\
0.08	0.0266082003234616\\
0.09	0.0389044657251616\\
0.1	0.0541247268196733\\
};
\addlegendentry{CSS [[13,1]] A = 100}

\end{axis}
\end{tikzpicture}%
	}%
	\caption{Performance of short codes over an asymmetric channel, $k=1$. Symmetric codes: $11$-qubits code with $t=2$ and $17$-qubits code with $t=3$. Asymmetric $13$-qubits code with $\eg = 1, \eZ=2$ and $15$-qubits code CSS with $t_{\mathrm X} = 1$ and $t_{\mathrm Z} = 3$.}
	\label{fig:PerformanceAsymCodesEZ2}
\end{figure}

About the results in Fig.~\ref{fig:PerformanceAsymCodes}, we remark that for symmetric codes the performance does not depend on the asymmetry parameter $A$, but just on the overall error probability $\rho$. For these codes, for a given $t$ the best \ac{CWEP} is obtained with the shortest code. 
As expected, the performance of the new asymmetric $[[9,1]]$ code improves as $A$ increases. 
In particular, for the symmetric channel, $A=1$, the $5$-qubits code performs better than the new one, due to its shorter codeword size. However, already with a small channel asymmetry, $A=3$, the new code performs better than the $5$-qubits code in the range of interest. For $A=10$ the new code performs similarly to the $[[11,1]]$ symmetric code with $t=2$. Asymptotically for large $A$, the channel errors tend to be of type $\PauliZ$ only, and consequently the new code %corrects up to $\eg+\eZ=2$ channel errors,
behaves like a code with $t=2$. % (note that a [[$9,1$]] code correcting $t=2$ generic errors does not exist). 
In the case of Shor code with extended error correction capability, it can be observed that when varying $A$ the curves are quite close each others. This is due that for the asymmetric channel the most important error patterns are the $\PauliZped{i}\PauliZped{j}$, and this code can correct only a subset of all possible patterns of this type. For the same reason the curve for $A=10$ is worse than that for $A=3$.
%
%%%%%%%%%%%%%%%%%%%%%%%%%%%%% eZ=2 %%%%%%%%%%%%%%%%%%%%%%%%%%%%%%%%%%
We next move to the case $\eg=1, \eZ=2$, and compare our asymmetric $[[13,1]]$ code of Table~\ref{tab:StabMatrixEG1K1EZ2} with two symmetric codes, and with the $[[15,1,3/7]]$ \ac{CSS} asymmetric code \cite{Sar:2009}. The performance of our asymmetric and of symmetric codes is given in general by \eqref{eq:PeGen}. For the \ac{CSS} code, we observe that it corrects all patterns with up to $\eg=1, \eZ=2$ errors, plus patterns with one $\PauliX$ and three $\PauliZ$ errors. Therefore, the \ac{CWEP} for the $[[15,1,3/7]]$ \ac{CSS} code is 
\begin{align}
\label{eq:PeCSS}
P_{\mathrm e} \:=&\: 1 - 15\binom{14}{3}\, \px \, \pz^3 \, \left(1-\rho\right)^{11} \nonumber \\
&- \sum_{j = 0}^{3} \binom{15}{j}(1-\rho)^{15-j}\, \xi(j;1)
\end{align}
where $\xi(\cdot;\cdot)$ is given by \eqref{eq:PeGenxi}. 
The analytical performance of the different \acp{QECC}, as given by \eqref{eq:PeGen} and \eqref{eq:PeCSS}, is reported in Fig.~\ref{fig:PerformanceAsymCodesEZ2}. We can see here that the proposed $[[13,1]]$  asymmetric code performs better than the \ac{CSS}, and provides a performance similar to the longer symmetric $[[17,1]]$ code for large $A$.
%%%%%%%%%%%%%%%%%%%%%%%%%%%%%%%%%%%%%%%%%%%%%%%%%%%%%
\section{Conclusions}
We have investigated a new class of stabilizer codes for quantum asymmetric Pauli channels, capable to correct up to $\eg$ generic errors, plus $\eZ$ errors of type $\PauliZ$. For these codes we generalized the quantum Hamming bound, and derived the analytical expression of the performance over asymmetric channels. We designed a $[[9,1]]$ \ac{QECC} which is the shortest, according to the new bound, capable to correct up to one generic error plus one  $\PauliZ$ error, and a $[[13,1]]$ \ac{QECC} capable to correct up to one generic error plus two $\PauliZ$ errors. The comparison with known symmetric \acp{QECC} confirms the advantage of the proposed codes in the presence of channel asymmetry.
%%%%%%%%%%%%%%%%%%%%%%%%%%%%%%%%%%%%%%%%%%%%%%%%%%%%%%%%%%%%%%%%%%%%%%%%%%%%%%%%%%%%%%%%

%

\bibliographystyle{IEEEtran}
\bibliography{IEEEabrv,File/Bibliography_QECC,File/quantumCU}

\begin{IEEEbiography}
%[{\includegraphics[width=1in,height=1.25in,clip,keepaspectratio]{BIOs/MarcoChiani.jpg}}]%keepaspectratio
{Marco Chiani}
(M'94-SM'02-F'11) received the Dr. Ing. degree ({\it summa cum laude}) in
electronic engineering and the Ph.D. degree in electronic and computer
engineering from the University of Bologna, Italy, in 1989 and 1993,
respectively.

He is a Full Professor of Telecommunications at the University of
Bologna. During Summer 2001, he was a Visiting Scientist at AT\&T
Research Laboratories, Middletown, NJ. Since 2003 he has been a frequent
visitor at the Massachusetts Institute of Technology (MIT), Cambridge,
where he presently holds a Research Affiliate appointment.
His research interests are in the areas of communications theory,
wireless systems, coding theory, and statistical signal processing.
In 2012 he has been appointed Distinguished Visiting Fellow of the Royal
Academy of Engineering, UK. He served as elected Chair (2002--2004) of the
Radio Communications Committee of the IEEE Communication Society and
as Editor (2000--2007) of Wireless Communication for the 
{\scshape{IEEE Transactions on Communications}}.

He received the 2011 IEEE Communications Society Leonard G. Abraham
Prize in the Field of Communications Systems, the 2012 IEEE
Communications Society Fred W. Ellersick Prize, and the 2012 IEEE
Communications Society Stephen O. Rice Prize in the Field of
Communications Theory.
\end{IEEEbiography}
\vspace{-0.8cm}
\begin{IEEEbiography}
%[{\includegraphics[width=1in,height=1.25in,clip,keepaspectratio]{BIOs/LorenzoValentini.jpg}}]
{Lorenzo Valentini} received the M.S. degree ({\it summa cum laude}) in Electronics and Telecommunications Engineering from the University of Bologna, Italy, in 2019. 
After graduation he has been with the Interdepartmental Centre for Industrial ICT Research of the University of Bologna, working on Internet of Things for the project CoACh. He is currently a Ph.D. student in Electronics, Telecommunications and Information Technologies Engineering at the University of Bologna. His research interests include communication theory, wireless sensor networks and quantum information theory.

\end{IEEEbiography}

\end{document}